\documentclass[structabstract]{aa}  
\usepackage{graphicx}
\usepackage{txfonts}
\usepackage{natbib}
\bibpunct{(}{)}{;}{a}{}{,} 
\usepackage{epsfig}
\usepackage{longtable}

\begin{document}
   \title{X-ray activity and rotation of the young stars in IC~348}

   \author{F.~Alexander\inst{1} \and T.~Preibisch\inst{1} 
          }

   \institute{Universit\"ats-Sternwarte M\"unchen, Ludwig-Maximilians-Universit\"at,
          Scheinerstr.~1, 81679 M\"unchen, Germany \email{frauke , preibisch@usm.uni-muenchen.de} }

   \date{Received September 15, 2011; accepted December 15, 2011}

 
  \abstract
   {The physical origin of the strong
    magnetic activity in T Tauri stars and its relation to stellar 
    rotation is not yet well-understood.
   }
   {We investigate the relation between the X-ray activity, rotation, and
    Rossby number
    for a sample of young stars in the $\approx 3$~Myr old cluster IC~348.
   }
   {We use the data of four \textit{Chandra} observations of IC~348 to derive the
    X-ray luminosities of the young stars. 
  Basic stellar parameters and rotation rates are collected from the
 literature. This results in a sample of 82 X-ray detected stars with known rotation 
  periods. 
 We determine the Rossby numbers (i.e.~the ratio of rotation period to convective 
turnover time) of 76 of these stars from stellar structure- and evolution-models
 for pre-main sequence stars.
   }
   {The young stars in IC~348 show no correlation between X-ray activity and
   rotation period. 
 For the Rossby numbers, nearly all IC~348 stars are in the 
  saturated regime of the activity--rotation relation defined by main-sequence stars.
Searching for possible super-saturation effects, we find a marginal (but statistically in-significant)
 trend that the stars with the 
smallest Rossby numbers have slightly lower X-ray activity levels.
There are no significant differences in the X-ray activity level for stars of different spectral 
 types and no relation between spectral type and Rossby number is seen. 
 In addition, for stars belonging to different IR-classes, no significant differences  are present
for the X-ray activity level as well as for their Rossby numbers. 
We compare the dispersion in the fractional X-ray luminosities of the stars in the saturated rotation regime 
in IC~348 to that seen in younger and older stellar populations.
The scatter seen in the  $\approx 3$~Myr old IC~348 
[$\sigma\left(\log\,(L_{\rm X}/L_{\rm bol}) \right) = 0.43$] is considerably smaller than
 for the $\approx 1$~Myr old Orion Nebula Cluster [\,$\sigma\left(\log (L_{\rm X}/L_{\rm bol}) \right) = 0.63$],
but, at the same time, considerably larger than the dispersion seen in the $\approx 30$~Myr old cluster NGC~2547 
[$\sigma\left(\log (L_{\rm X}/L_{\rm bol}) \right) = 0.24$] and in
main-sequence stars.
   }
   {
The results of our X-ray analysis of IC~348 show that neither the rotation rates nor the presence/absence of 
circumstellar disks are of fundamental importance for
determining the level of X-ray activity in TTS.
Our results suggest that the scatter in the X-ray activity levels for the rapidly rotating
members of young clusters
decreases with the age of the  stellar population. 
We interpret this
as a signature of the changing interior structure of pre-main sequence stars
and the consequent changes in the dynamo mechanisms that are responsible for the magnetic field generation.}
   \maketitle
%

\section{Introduction}

Young stellar objects (YSOs) in all evolutionary stages, from 
class~I protostars, to T Tauri stars (TTS) to zero-age main-sequence 
stars, have highly ($\sim 10^3 - 10^4$ times) elevated levels of X-ray 
activity (for reviews of the X-ray properties of YSOs
and stellar coronal astronomy in general, see Feigelson \& Montmerle 1999 and
Favata \& Micela 2003).
Despite many years of research, the
physical origin of this X-ray activity remains poorly understood.
Although there is strong evidence that in most TTS
the X-ray emission originates from magnetically confined coronal plasma 
\citep[e.g.][]{Preibisch05},
it is unclear what kind of 
dynamo processes create the required magnetic fields.

For main-sequence stars, the level of the magnetic activity (and thus the
strength of the X-ray emission) is mainly determined by
their rotation rate. Observations have revealed a 
clear rotation--activity relation
of the form  $L_{\rm X}/L_{\rm bol} \propto P_{\rm rot}^{-2.6}$
\citep[e.g.][]{Pallavicini81,Pizzolato03},
which is in good agreement with the expectations of
 solar-like $\alpha\!-\!\Omega$ dynamo models 
\citep[e.g.][]{Maggio87,Ossendrijver03}.
The solar dynamo is thought to be anchored in the
``tachocline'', a thin zone between the inner radiative core and
the outer convection zone. However, 
the connection between the observed surface magnetic activity and 
the properties of the dynamo generating the magnetic flux is not yet
fully understood \citep[e.g.,][]{Isik11}. 

For main-sequence stars,
the increase in magnetic activity towards shorter rotation periods
stops for periods shorter than $\sim 2-3$ days, where the activity saturates 
around $\log\left(L_{\rm X}/L_{\rm bol}\right) \approx -3$.
The physical reasons for this saturation effect are poorly understood
\citep[see, e.g.,][]{Jardine99}.
For extremely rapidly rotating stars, the activity level seems to
decrease slightly with increasing rotation rate \citep{Prosser96,Randich00,James00,Jeffries11},
a phenomenon that is denoted as ``super-saturation'' and also not well-understood.

Studies of stellar
populations with different ages show that that there is a continuous
evolution from the very high X-ray activity levels in the youngest stages
to the much lower activity seen in older (more than a few hundred Myr old) stars \citep[e.g.,][]{Guedel97,PF05,Telleschi05}.
This evolution can be explained by the temporal decrease in the stellar angular momentum
\citep{Bouvier97,Herbst07}.
Furthermore, the presence of strong magnetic fields on the surface
of T Tauri stars has been clearly established \citep[e.g.][]{Johns-Krull07}.
These pieces of evidence suggest that 
the X-ray activity of YSOs originates in dynamo processes similar 
to those present in our Sun.

However, the relation between rotation and X-ray activity 
in TTS remained unclear until recently, since
in most studies of star forming regions the number of
X-ray detected TTS with known rotation periods was too small
to draw sound conclusions.

A few years ago,
the $Chandra$ Orion Ultradeep Project (COUP),
a ten-day long observation of the Orion Nebula Cluster (ONC)
with $Chandra$/ACIS \citep[for details of the observation and
data analysis, see][]{Getman05} 
and the XMM-Newton Extended Survey of the Taurus Molecular Cloud 
\citep[XEST, see][]{Guedel07}
provided very sensitive X-ray data sets for large samples of
TTS.
The COUP and the XEST data have both shown 
that the TTS in Orion and Taurus do \textit{not} follow
the relation between rotation period and X-ray luminosity for main-sequence stars
\citep{Preibisch05,Briggs07}. In addition the TTS spin up during the first $\sim$10-30 Myr \citep{Herbst05} does not lead to an increase in the X-ray luminosity.
This places doubt on the solar-like
dynamo activity scenario for TTS.
Another argument against solar-like dynamos in young TTS comes from theoretical
considerations: at ages of  $\le 2$~Myr, most TTS are expected to be
fully convective, and thus should not possess a tachocline.
The conventional $\alpha\!-\!\Omega$ dynamo cannot work in such 
a situation.
Several alternative dynamo concepts have been developed
for fully convective stars
\citep[e.g.][]{Giampapa96,Kueker99,Dobler06,Chabrier06,Browning07,Voegler07,Graham10}.
Although the reliability of these theoretical models is not entirely clear,
there is good evidence for the simultaneous presence
of an $\alpha\!-\!\Omega$ dynamo \textit{and} some kind of a small-scale turbulent dynamo in the
convection zone of our Sun \citep[e.g.,][]{Durney93,Bueno04}.

Direct observations of the 
relation between rotation and magnetic activity for stellar samples
spanning a wide range of ages can provide 
fundamental constraints on the theoretical models.
Numerous large datasets are available for samples of main-sequence stars
as well as for young stellar clusters with ages
as young as $30$~Myr \citep[e.g.][]{Prosser96,Stauffer97a,Stauffer97b,Randich00,Jeffries11}.
For younger ages, however, there is still a clear lack of 
reasonably large stellar samples for which 
good activity \textit{and} rotation data are available.
The data obtained in the COUP and XEST observations 
do both indicate that there are
very young stellar populations of only
$\la 1$~Myr old \citep[see, e.g.,][]{Weights09,Dib10,Luhman10}.
Since the stellar rotation, the magnetic activity levels, and other
basic stellar parameters
evolve strongly in the age range between 1~Myr and 30~Myr,
a sample of stars with an age of a few Myr can provide valuable information.
With an age of $\approx 3$~Myr \citep{Luhman03,Mayne07}, the young cluster 
IC~348 is very well-suited in this respect.
This age is particularly interesting because it corresponds to the point in 
time when the structure of solar-mass stars changes from a fully convective 
interior to a radiative core plus convective envelope structure, 
and this should affect the dynamo processes that are the ultimate source of the magnetic activity.

\section{X-ray and rotation data for IC~348}

IC\,348 is the nearest ($\approx 310$\,pc, \citet{Herbig98}) rich and compact
very young stellar cluster \citep{Herbst08}.
In a large number of observational studies, more than 300 individual
cluster members
have been identified and accurately characterized using optical and infrared
spectroscopy and photometry
\citep{Herbig54, Herbig98, Lada95, Lada98,
Muench03, Preibisch03,Luhman98,Luhman99,Luhman03,Luhman05}.
For nearly all of these stars, basic parameters
such as luminosity, effective temperature, mass, and age are known.
The mean age of the cluster members is $\approx 3$\,Myr. 
Extensive studies with the \textit{Spitzer} Space Telescope have provided comprehensive information about the
frequency and nature of the circumstellar disk population
in IC~348 
\citep{Lada06, Muench07, Currie09}.

\subsection{\textit{Chandra} X-ray observations}

We targeted IC\,348 during four different observations  with the Imaging Array of the \textit{Chandra} Advanced 
CCD Imaging Spectrometer (ACIS-I). The first observation was obtained in September 2000 
(ObsID 606, 52.9~ksec exposure time, PI: Th.~Preibisch), and the results were presented in 
\citet{PZ01} and \citet{PZ02}. Three additional observations were performed in 
March 2008 (ObsID 8584, 50.1~ksec exposure time, PI: N.~Calvet, 
ObsID 8933, 40.1~ksec exposure time, PI: S.~Wolk, and ObsID~8944, 38.6~ksec exposure time, PI: S.~Wolk).
While ObsID 606 was centered at the optical cluster center, ObsID 8584 is shifted 
by $5.6'$ to the south-west, and ObsIDs~8944 and 8944 are shifted  $13.1'$ to the south-west.
The total area covered by the combined \textit{Chandra} data set is about 555 square-arcmin
and includes most known cluster members.

For the study presented in this paper, we merged and analyzed all four 
\textit{Chandra} data sets to determine the X-ray luminosities of the TTS
in IC~348.
The source detection in the merged image
was performed in a standard way by using the \textit{wavdetect} algorithm
in \textit{CIAO} to locate X-ray sources in our merged image. The resulting preliminary source
list  was extended by adding additional possible sources identified by visual inspection.
This yielded a catalog of 392 possible X-ray sources.
To clean this catalog of spurious sources and determine the properties of the X-ray
sources, we performed a detailed analysis of each individual candidate source with the 
\textit{ACIS Extract} software 
package\footnote{http://www.astro.psu.edu/xray/docs/TARA/ae\_users\_guide.html} \citep{Broos10},
following the procedures described in \citet{Getman05}, \citet{Townsley03}, and \citet{Broos07}.
The final catalog contains 290 X-ray sources.

Intrinsic, i.e. extinction-corrected, X-ray luminosities for the sources were determined using the \textit{XPHOT} 
software\footnote{http://www.astro.psu.edu/users/gkosta/XPHOT/}, developed by \citet{Getman10}, assuming a distance of 310$\,pc$.
For sources with fewer than four net counts (for which \textit{XPHOT} does not work), 
an estimate of the X-ray luminosity was derived from the $FLUX2$ values computed by 
\textit{ACIS Extract} 
and the median energy of the detected photons.
For those cluster members that were not detected as X-ray sources in the \textit{Chandra} data,
we determined upper limits to their X-ray luminosities in the following way.
First, extraction regions were defined as the 90\% contours of the local PSF and then 
the  number of counts in the target aperture and an estimate of the local background
was determined with \textit{ACIS Extract}. Upper limits to the number of source counts
were then calculated with the Bayesian method to determine 
confidence intervals involving Poisson-distributed data described in \citet{Kraft91},
using a confidence level of 0.9. To obtain upper limits to the X-ray luminosity, these source count upper limits were devided by the
local exposure time and multiplied with the mean conversion factor between the count rate and the X-ray luminosity
derived 
for the X-ray detected TTS from our IC~348 sample.

A general description of the resulting X-ray properties and an investigation of relations
to the basic stellar parameters is given in \citet{Stelzer11}.

\subsection{Rotation periods of IC~348 stars from the literature}

Numerous determinations of rotation periods have been performed in the past few years for stars in IC~348 starting with \citet{Herbst2000}.
From the studies of \citet{Cohen04}, \citet{Littlefair05}, \citet{Kiziloglu05}, \citet{Cieza06}, and \citet{Nordhagen06},  
we collected a catalog with rotational periods for 206 stars in IC~348.
For most stars that are listed in more than one of these studies,
the rotation periods agreed to within typically a few percent.
We used the mean value of the listed rotation periods in these cases.
In a few cases of significant discrepancies, we
 calculated the mean after dropping the outlying value.

\subsection{Construction of the rotation - activity catalog}
The starting point for the construction of the rotation-activity catalog was the membership 
list for IC~348 from \citet{Luhman03}, which contains stellar parameters such as
 bolometric luminosity, spectral type, and effective temperature for 288 members. 
This sample is thought to be complete for $A_{V} \le 4$~mag and masses of 
$M \ge 0.03 M_{\odot}$.
For each star with known rotation period, we searched for an
associated X-ray source within $\approx 2''$ of the catalogued 
optical position.
For the 18 cluster members with known rotation periods that are
located in the \textit{Chandra} area but not detected as X-ray sources, we
used the upper limits to the X-ray luminosity as described above whereas stellar parameters were taken from \citet{Luhman03} and \citet{Muench07}.

Our final rotation - activity catalog consists of 82 stars
with measured X-ray luminosities and 18 stars with upper limits to the
X-ray luminosity.
Tables \ref{tab:parameters} and \ref{tab:upper_limits} list the stellar parameters and X-ray luminosities for the
100 stars in our final sample.

\section{X-ray activity and rotation periods}

  \begin{figure}
  \includegraphics[width=9cm]{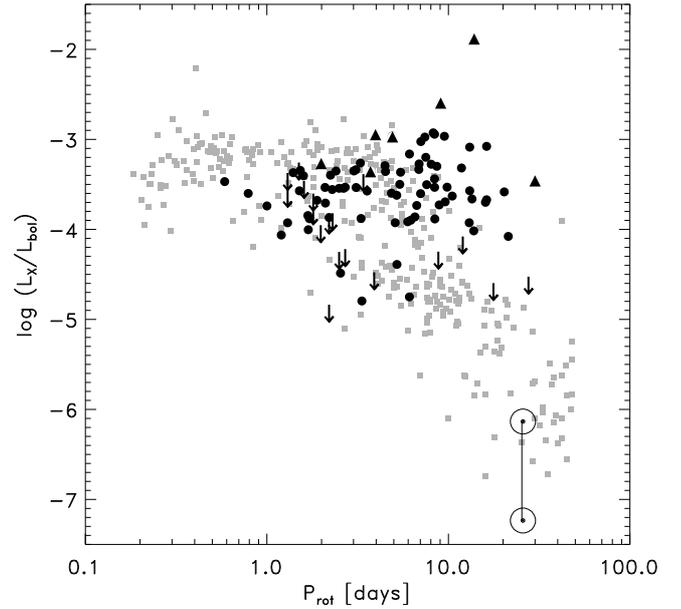}
  \caption{Fractional X-ray luminosity versus rotation period.
This plot compares the IC348 TTS (solid black dots)
to MS stars from \cite{Pizzolato03} and \cite{Messina03}
(gray boxes) and the Sun. IC~348 sources with strong flares during the \textit{Chandra} observation are marked by triangles.
\label{lxlb-prot:fig}}
    \end{figure}

In Fig.~\ref{lxlb-prot:fig}, we show the fractional X-ray luminosities 
$L_{\rm X}/L_{\rm bol}$ versus 
rotation periods for the TTS in IC~348 and compare them to data for main-sequence
stars. 
The periods for the stars in our sample range from $0.585$~days to $30$~days.
It is obvious that the IC~348 stars do not follow the 
well-established activity-rotation relation defined by the main-sequence stars,
i.e.~increasing activity for decreasing
rotation periods followed by saturation at $L_{\rm X}/L_{\rm bol} \sim 10^{-3}$
for periods shorter than about three days.
Instead, the IC~348 stars show no relation between X-ray activity and 
rotation period. To identify stars that displayed large X-ray flares during the Chandra observation, we inspected the X-ray lightcurves. Stars for which the amplitude of the count rate variation in the lightcurve is higher than 10 are marked by triangles in Fig.~\ref{lxlb-prot:fig}. 
We note that the two stars with the highest fractional X-ray luminosities ($\log \left(L_{\rm X}/L_{\rm bol}\right)
=-2.6$ for star J\,034427.02+320443.6 and 
$\log\left(L_{\rm X}/L_{\rm bol}\right)=-1.9$ for star J\,034359.69+321402.9)
showed strong X-ray flares during the \textit{Chandra} observation; 
the amplitude of the count rate variation was 29.6 and 36.5, respectively, and the exponential decay time of the 
flares was $\sim$ 4-6~ksec. 
Therefore, the effect of these short flares on the mean values of X-ray luminosity
averaged over the entire duration of the observation is small.
\footnote{A detailed analysis of the effects of flares
on the mean X-ray luminosity X-ray of TTS was performed
in the context of the
\textit{Chandra} Orion Ultradeep Projects \citep{Getman05}.
An analysis of the X-ray lightcurves
was used to derive the ``characteristic'' X-ray luminosities by
removing the effect of strong flares. It was found that the
difference between these ``characteristic'' X-ray luminosities
and the average X-ray luminosities (i.e.~the mean values determined
without removing flares) is very small and does not
significantly affect the relations
between X-ray and basic stellar properties \citep{Preibisch05}.
Our IC~348 \textit{Chandra} observations have long enough
exposure times (compared to the typical short duration
of a flare) for individual flares to have only very minor effects on the average X-ray luminosities.}

The rotation periods of the X-ray undetected stars are rather uniformly
distributed across the full range of periods for the X-ray detected stars.
This implies that the X-ray detection limit does not produce a systematic
bias against either very rapid or slow rotators.

The data show that the X-ray activity level for the TTS in IC~348 seems to be
independent of the rotation period. 
For the rapid rotators (periods $\leq 5$~days), the absence of an activity--rotation
correlation may be an effect of the saturation seen in main-sequence stars.
However, there is a considerable
number of rather slowly rotating IC~348 stars (periods $\ga 10$~days) that
have much higher X-ray activity levels than similarly slowly rotating main-sequence stars.
We can thus confidently conclude that the IC~348 TTS do not follow the same
relation between rotation period and fractional X-ray luminosity as seen for main-sequence stars.

\section{X-ray activity and Rossby numbers}

Although the efficiency of the magnetic field generation in the $\alpha\!-\!\Omega$ 
dynamo model increases with the rotation
rate, there is no direct causal relationship. The
dynamo number depends instead on the radial gradient of the angular velocity and the
characteristic scale length of convection  at the base of the
convection zone \citep[see][]{Ossendrijver03}. A detailed theoretical analysis shows that 
the dynamo number scales as the inverse square of the
Rossby number $Ro$
\citep[e.g.][]{Maggio87}, which is
defined as the ratio of the rotation period to the
convective turnover time $\tau_c$, i.e.~$Ro := P_{\rm rot}/\tau_c$.
This theoretical prediction is well-confirmed by data for
main-sequence stars, which follow a considerably tighter relationship
between magnetic activity and Rossby number than magnetic activity and rotation period
\citep[e.g.][]{Montesinos01,Pizzolato03}.
Numerous studies have confirmed that for slowly rotating stars,
activity rises as $L_{\rm X}/L_{\rm bol} \propto Ro^{-2}$.
Saturation at the $L_{\rm X}/L_{\rm bol} \approx 10^{-3}$ level
is reached around $Ro \sim 0.1$,
which is followed by a regime of ``supersaturation''
for very small Rossby numbers, $Ro \lesssim 0.02$ \citep[e.g.,][]{Randich00,Jeffries11}.

\subsection{Determination of convective turnover times for the TTS}

The convective turnover timescale $\tau_c$ is a sensitive function of
the physical properties in the stellar interior. For main-sequence stars,
semi-empirical interpolations of $\tau_c$ values as a function of the
color or the stellar mass are available and provide an easy way to
estimate the Rossby numbers.
For TTS, which have a considerably different and
rapidly evolving internal structure, the situation is far more
complex because the $\tau_c$ values depend strongly on age.
The convective turnover timescales for the young TTS 
are several times longer than for main-sequence stars of the same mass.
Detailed stellar evolution model calculations are required to 
determine reliable convective turnover times for TTS in a self-consistent way.

Model calculations of this type were performed for the analysis
of the activity -- rotation relation in the Orion Nebula 
Cluster described in \citet{Preibisch05}. First, a large set of 
stellar models was computed with the
Yale Stellar Evolution Code \citep[see][]{Kim96}
for stellar masses between $0.065\,M_\odot$ and
 $4.0\,M_\odot$. In the second step,
the convective turnover time for each star in the observed sample
was determined 
from the model that reproduced its observed effective temperature and
luminosity.
This resulted in a database of 562 different
$\left(T_{\rm eff}, L_{\rm bol}, \tau_c \right)$ values.

We used these results to determine convective turnover times
for the TTS in IC~348 by interpolating their stellar parameters
$T_{\rm eff}$ and $L_{\rm bol}$ against those in the database
to determine $\tau_c$. This resulted in Rossby numbers for 76 stars in our sample.

\subsection{The activity -- Rossby number relation for the TTS}

  \begin{figure}
  \includegraphics[width=9cm]{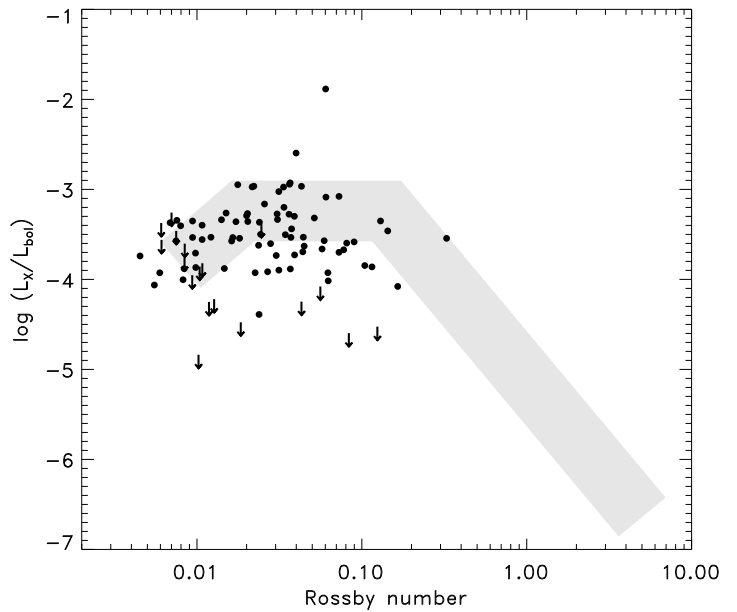}
  \caption{
Fractional X-ray luminosity versus Rossby number
for the IC~348 stars. Black arrows indicate upper limits on the X-ray activity.
The gray shaded area shows the relation and the width of its
 typical scatter found for MS stars \citep[from][]{Pizzolato03}.
\label{lxlb-rossby1:fig}}
    \end{figure}

%

Rossby numbers were computed by dividing the
rotation periods of the stars
by the values for their local convective turnover time derived above.
Fig.~\ref{lxlb-rossby1:fig} shows the
fractional X-ray luminosities
of the TTS versus the resulting Rossby numbers.
The plot shows no strong relation between these two quantities as expected for such low Rossby numbers.
Nearly all TTS have Rossby numbers $< 0.2$ and are therefore
in the saturated or super-saturated
regime of the activity -- Rossby number relation for main-sequence
 stars.

To search for indications of super-saturation,
we compared the fractional X-ray luminosities
of the TTS in the saturated ($0.1 > Ro > 0.02$) and super-saturated
($Ro \leq 0.02$) regimes. We found nearly equal median values of
$\log\left(L_{\rm X}/L_{\rm bol}\right)=-3.51$ for the TTS
in the saturated regime and $-3.56$ for
those in the super-saturated regime; thus, the stars from IC~348 show no clear evidence of supersaturation.
Nevertheless, there may be a hint of supersaturation  if we consider
only stars with very small Rossby numbers of $Ro \leq 0.006$.

All three stars in this range have remarkably low activity levels of between
$-3.7$ and $-4.1$.
If this is a real effect, 
the border for supersaturation would be shifted by about a factor of three relative to the 
border determined by \citet{Randich98} for main-sequence stars.

A difference between TTS and main-sequence stars manifests itself 
in the wide dispersion of fractional X-ray luminosities at a given Rossby number among the TTS. 
The activity levels scatter over more than one order of magnitude,
in remarkable contrast to the much smaller scatter found
for main-sequence stars in the saturated rotation regime \citep[see][]{Pizzolato03}.
This result suggests that additional factors, other than the rotation period, are important for 
determining the level of X-ray activity in TTS.

\subsection{The activity -- Rossby number relation for stars of different spectral types and infrared-classes}

\begin{figure}
  \includegraphics[width=9cm]{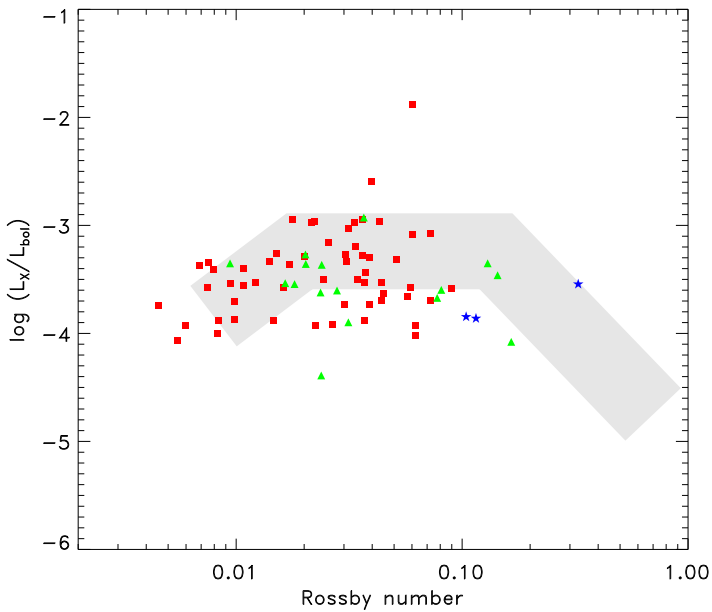}
  \caption{
Comparison of fractional X-ray luminosity versus Rossby number
for stars in IC~348 for the different spectral types G (blue asterisks), K (green triangles), and M (red squares).
The gray shaded area shows the relation and the width of its
 typical scatter found for MS stars \citep[from][]{Pizzolato03}.
\label{lxlb-rossby3:fig}}
    \end{figure}

A comparison of $L_{\rm X}/L_{\rm bol}$ versus Rossby numbers for different spectral classes
is shown in Fig.~\ref{lxlb-rossby3:fig}. 
 
We found no statistically significant difference between the median values of the fractional luminosity for the different spectral types.
In addition, the ranges
of Rossby numbers for the different spectral types seem to be quite similar.
The three G-type stars have larger Rossby numbers than the K- and M-type stars,
although this may simply be an effect of small-number statistics.
%

\begin{figure}
  \includegraphics[width=9.0cm]{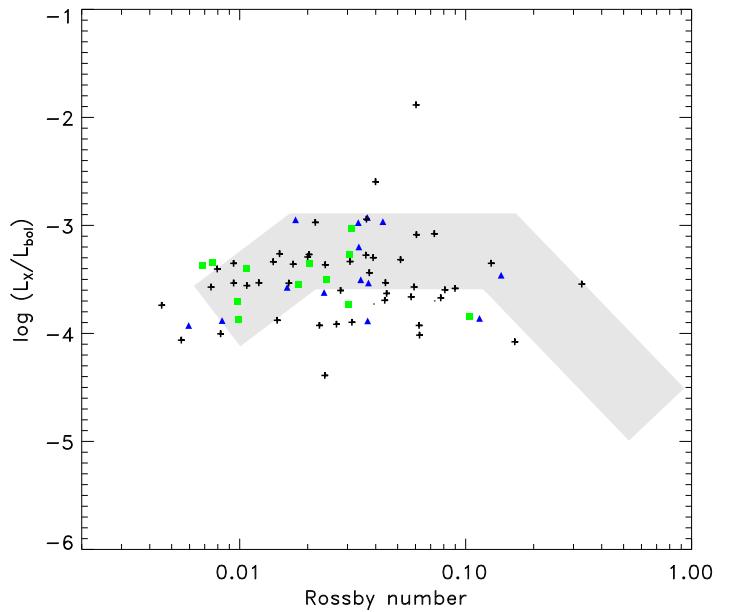}
  \caption{
Comparison of fractional X-ray luminosity versus Rossby number
for stars in IC~348 for the different IR-classes II (blue triangles), II/III (green squares), and III (black crosses).
The gray shaded area shows the relation and the width of its
 typical scatter found for MS stars \citep[from][]{Pizzolato03}.
\label{lxlb-rossby4:fig}}
    \end{figure}

In Fig.~\ref{lxlb-rossby4:fig}, we show X-ray activity versus Rossby number for the 
different infrared-classes, which trace the circumstellar material around the young stars.
We used the infrared-classes derived by \citet{Lada06} from the observed \textit{Spitzer}/IRAC
spectral energy distributions (SEDs) between $3.6\,\mu$m and $8\,\mu$m.
Objects with an SED slope of $\alpha_{\rm 3-8} >-0.5$ are class~1 and are thought
to be very young stellar objects surrounded by circumstellar disks and 
envelopes.
Class\,II objects ($-0.5 \geq \alpha_{\rm 3-8} \geq -1.8$) are stars with
thick circumstellar disks. Class\,II/III objects ($-1.8 \geq \alpha_{\rm 3-8} \geq -2.56$) 
are thought to be stars surrounded by ``anemic'' disks, whereas class\,III objects
($\alpha_{\rm 3-8} < -2.56$) are disk-less stars. 
The ``anemic'' disk objects are probably transition objects that have
either  optically thin disks, or disks with large inner holes \citep[see][]{Lada06}. 

The plot shows no obvious differences in either the X-ray activity levels 
or the Rossby numbers of the different infrared classes.
This suggests that the presence or absence and the properties of the circumstellar
material are independent of the level of X-ray emission.
This is consistent with the results derived from \textit{Chandra} observations
of the Orion Nebula Cluster \citep{Preibisch05}.

Investigating the plot of rotation periods versus fractional luminosities for the different infrared classes, we found no significant relation. 
We also found no indications of a relation between the Rossby number and the infrared class.

\section{Comparison of IC~348 to young clusters of different age}

During the pre-main sequence lifetime of young stars, several
important changes occur with respect to the level and origin of the
magnetic activity. 
X-ray studies of young clusters with different age have shown that the
level of X-ray activity is approximately constant 
during the initial  $\sim 10$~Myrs, and starts to decline significantly
at ages above $\sim 10-30$~Myr \citep{PF05}. 
Another important aspect is that the internal structure of the young stars
changes from a fully convective structure to a radiative core plus
convective envelope structure. The timescale for these transitions depends
on the stellar mass; for 
stars with masses of around one solar mass, this change occurs at ages around $2-4$~Myr. 
It is therefore interesting
to compare the rotation--activity relation for IC~348 to that seen in 
other young clusters of different age.
While there is certainly no lack of X-ray data on young stellar clusters,
rotation data are often unavailable for a sufficiently large number of
cluster members for statistical studies, especially for the particularly
interesting age period of $\le 10$~Myr.
 Another serious obstacle is often
the non-availability of reliable Rossby numbers for young pre-main sequence stars.

Here we consider two other young clusters for which similarly good
X-ray and rotation data as well as reliable Rossby numbers are available
for a sufficiently large number of stars.
The first data set is the sample of TTS in the $\approx 1$~Myr old Orion Nebula
Cluster from \citet{Preibisch05}.
The second sample is taken from the recent study of the $\approx 30$~Myr old cluster
NGC~2547 by \citet{Jeffries11}.

\begin{figure*}
   \centering
   \includegraphics[width=9.0cm]{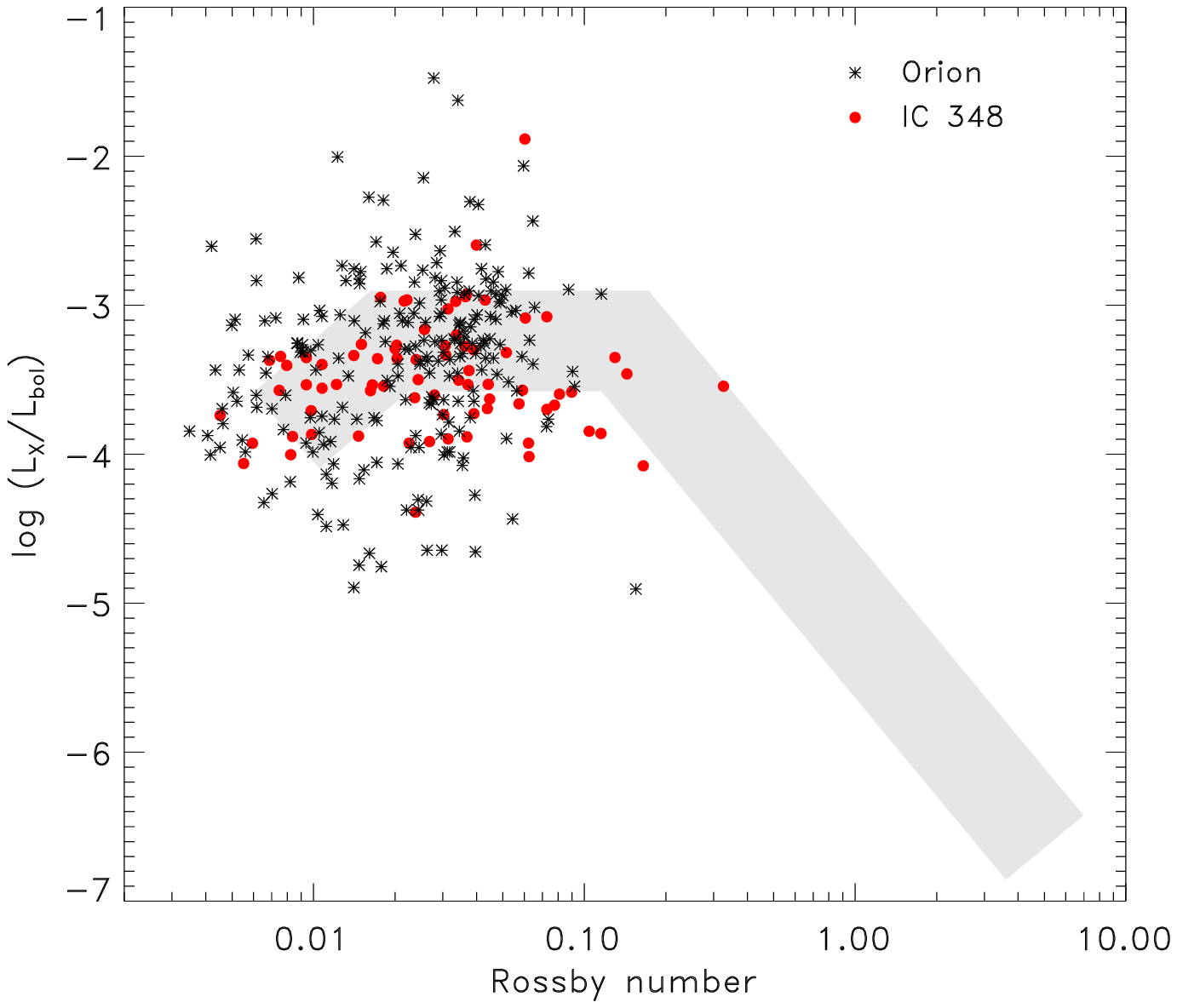} \
   \includegraphics[width=9.0cm]{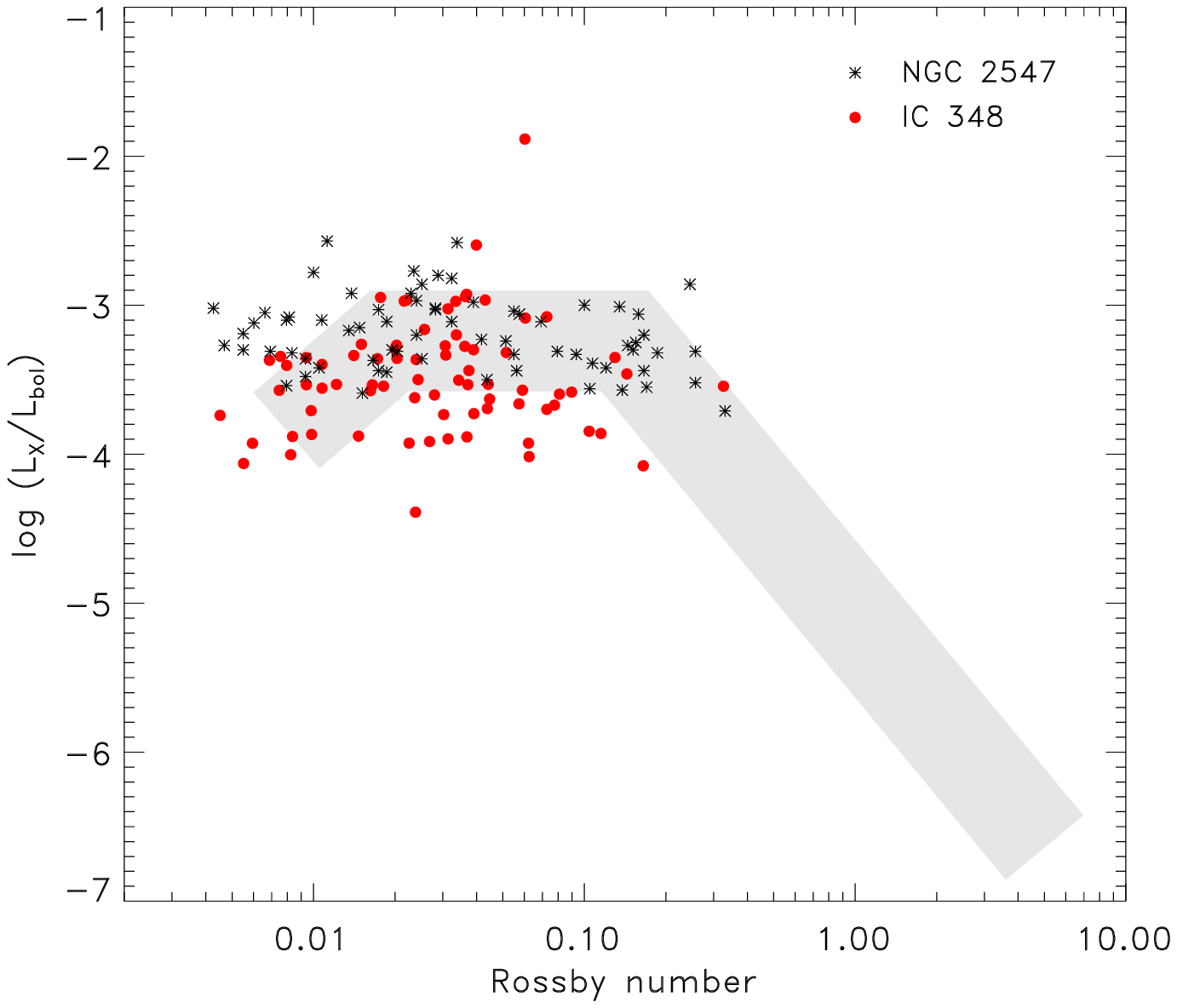}\
   \caption{Fractional X-ray luminosity versus Rossby number
for stars in IC~348 (red filled circles) compared to the stars (black asterisks) in the Orion Nebula Cluster (left) 
and NGC~2547 (right).
The gray shaded area shows the relation and the width of its
 typical scatter found for MS stars \citep[from][]{Pizzolato03}.
   }
              \label{lxlb-rossby2:fig}%
    \end{figure*}

In Fig.~\ref{lxlb-rossby2:fig}, we compare the relations between X-ray activity and
Rossby numbers for the three studied clusters.
The three samples are similar in the sense that no correlation can be seen between 
X-ray activity and Rossby number. A striking difference is however seen in the
scatter in the X-ray activity levels for the stars in the saturated Rossby number regime.
Table \ref{tab:statistics} lists statistical values for the scatter in the
activity levels seen in each sample.
The scatter is largest in the Orion Nebula Cluster sample
[\,$\sigma\left(\log (L_{\rm X}/L_{\rm bol}) \right) = 0.63$], but
considerably smaller in IC~348 [\,$\sigma\left(\log (L_{\rm X}/L_{\rm bol}) \right) = 0.43$], 
and yet smaller in the case of NGC~2547 [\,$\sigma\left(\log (L_{\rm X}/L_{\rm bol}) \right) = 0.24$].
We note that the scatter seen in NGC~2547 is representative of the scatter seen in main-sequence stars
\citep[see][]{Pizzolato03}.
As the age of the Orion Nebula Cluster sample is $\approx 1$~Myr, whereas IC~348
is $\approx 3$~Myr old, and NGC~2547 is $\approx 30$~Myr old, we find a decrease in the
scatter in the X-ray activity levels for stars in the saturated Rossby number regime
with the age of the stellar population.

\begin{table}
\caption{Statistical values (mean, standard deviation, median, and median absolute deviation)
 for the activity level $\log\,(L_{\rm X}/L_{\rm bol})$ of young stars in the Orion Nebula Cluster, 
IC~348, and NGC~2547.}
\label{tab:statistics}
\centering
\begin{tabular}{l r r r r }
\hline\hline
\noalign{\smallskip}
{Stat. value}&{Orion}&{IC~348}&{NGC~2547}\\
\noalign{\smallskip}
\hline
\noalign{\smallskip}
   {Mean}               &$-3.39$&$-3.53$&$-3.20$\\	 
   {Std. Dev.}          &0.63   &0.43   &0.24 \\
   {Median}             &$-3.31$&$-3.53$&$-3.23$\\
   {MAD}                &0.39   &0.21   &0.18 \\

\hline                                   
\end{tabular}
\end{table}

\section{Conclusions}

Our analysis of the relation between X-ray activity (as traced by deep \textit{Chandra}
observations) and the rotation properties of the TTS in the young cluster IC~348
has yielded the following results.
First, we have shown that there is no correlation between the fractional X-ray luminosity
and the rotation periods of the TTS; even the rather slowly rotating stars
(periods $\ga 10$~days) display very strong X-ray activity.
Second, according to the Rossby numbers (that we have determined based on
detailed stellar models for pre-main sequence stars), all TTS are in the
saturated regime of the rotation--activity relation defined by main-sequence stars.
Third, we have found no significant evidence of super-saturation among the most rapid
rotators, although our data suggest that stars with extremely low Rossby numbers
($Ro \leq 0.006$) have slightly lower activity levels.
This seems to agree with the results of \citet{Jeffries11}, who claim that 
in their NGC~2547 sample supersaturation occurs only for Rossby numbers lower than 0.005. 
In all three of these aspects, the TTS in IC~348 behave similarly to the TTS in 
the Orion Nebula Cluster.

A remarkable difference is however seen in the scatter in the X-ray activity
levels for the TTS in the saturated rotation regime.
The scatter seen in the $\approx 3$~Myr old IC~348 sample is considerably smaller than that in the
$\approx 1$~Myr old Orion Nebula Cluster sample, but, at the same time, considerably larger than seen 
in the $\approx 30$~Myr old stars in NGC~2547 as well as for main-sequence stars.
This suggests that some process reduces the wide distribution of activity levels seen in the
youngest stars towards the much narrower distribution in older pre-main sequence 
and (young) main sequence stars during the first $\sim 30$~Myr period, whereas the
absolute level of the X-ray activity remains roughly constant during that time.

A possible explanation of this effect may be related to the stellar interior structure 
and the corresponding dynamo mechanisms that are the basis of the magnetic activity
that produces the observed X-ray emission.
At an age of $\leq 1$~Myr (e.g., in the Orion Nebula Cluster), 
almost all low- and intermediate mass stars ($M \leq 2\,M_\odot$) are fully convective.
As already mentioned, the solar-like $\alpha\!-\!\Omega$ dynamo cannot operate in such a
situation and some kind of small-scale convective dynamo must be the source of the
magnetic activity. The more or less chaotic nature of such a dynamo may be 
responsible for the very wide scatter in activity levels seen for stars with 
similar rotation rates and stellar parameters.
According to the pre-main sequence stellar evolution models of \citet{Siess00}, 
a $1.5\,M_\odot$ star develops a radiative
core at an age of about 1~Myr.
As time proceeds, stars with increasingly lower masses develop radiative cores
(at $\approx 1.7$~Myr for $1.2\,M_\odot$ stars, at $\approx 2.3$~Myr for $1.0\,M_\odot$ stars, and
at $\approx 4.2$~Myr for $0.8\,M_\odot$ stars).
As soon as a star has a radiative core,
the conditions for the operation of a solar-like $\alpha\!-\!\Omega$ dynamo are met,
which then gradually replaces the convective dynamo.
When the total magnetic activity is dominated by the $\alpha\!-\!\Omega$ dynamo,
stars in the saturated rotation regime have a rather homogeneous level of
X-ray activity, i.e.~a small scatter in their $L_{\rm X}/L_{\rm bol}$ values.

At the 3~Myr age of IC~348, all stars with $M \ge 0.9\,M_\odot$ (corresponding to 
a spectral type of K6) should have a radiative core; this concerns about 25\% of our stellar sample in IC~348 and can explain why the dispersion in the X-ray activity levels in IC~348 is smaller than in the Orion Nebula Cluster.
As time proceeds, an increasing fraction of the lower-mass stars make the 
transition from a fully convective to a core/envelope structure, where
the operation of an $\alpha\!-\!\Omega$ dynamo becomes possible, and
this should continuously decrease the scatter in their $L_{\rm X}/L_{\rm bol}$ values.
By the age of $\approx 30$~Myr, nearly all stars have attained
their final stellar structure, and thus the scatter in the activity levels
has settled to the rather small dispersion as typical of (rapidly rotating) main-sequence stars.
This scenario could be a qualitative explanation of the
observed decrease in the scatter in the X-ray activity level
of stars in the saturated rotation regime.

\begin{acknowledgements}
We would like to thank the LMU student Stefan Heigl for his help in collecting
the literature data.
We gratefully acknowledge support for this project
 from the Munich
Cluster of Excellence: ``Origin and Structure of the Universe''.
\end{acknowledgements}

\longtab{2}{
\begin{longtable}{c c c c c c c c}
\caption{\label{tab:parameters} Stellar parameters for the sample of stars in IC~348}\\
\hline\hline
Ra, Dec&$L_{\rm bol}$&Spectral Type&Turnover time&Period&Class&X-ray Luminosity\\
 &$L_{\rm \odot}$&&days&days&&erg/s \\
\hline
\endfirsthead
\caption{continued.}\\
\hline\hline
Ra, Dec&$L_{\rm bol}$&Spectral type&Turnover time&Period&Class&X-ray luminosity\\
 &$L_{\rm \odot}$&&days&days&&erg/s \\
\hline
\endhead
\hline
\endfoot
  034424.57+320357.1	& 0.44  &    M1	  & 227.7 &    4.92  &      III  &  1.80$\cdot10^{30}$  \\                                         
  034422.57+320153.7	& 0.51  &  M2.5	  & 221.4 &    1.00  &      III  &  3.57$\cdot10^{29}$  \\                                         
  034421.26+320502.4	& 0.32  &  M2.5	  & 223.2 &    6.87  &      III  &  5.70$\cdot10^{29}$  \\                                         
  034422.29+320542.7	& 0.54  &    K8	  & 209.0 &   30.00  &       II  &  7.18$\cdot10^{29}$  \\                                         
  034421.66+320624.8	& 0.29  & M2.75	  & 223.7 &    8.38  &      III  &  4.06$\cdot10^{29}$  \\                                         
  034411.26+320612.1	& 0.39  &    M0	  & 234.9 &   10.50  &      III  &  3.52$\cdot10^{29}$  \\                                         
  034420.02+320645.5	& 0.12  &  M3.5	  & 221.6 &    8.63  &      III  &  2.32$\cdot10^{29}$  \\                                         
  034406.79+320754.1	& 0.17  & M4.25	  & 218.6 &    1.30  &       II  &  7.74$\cdot10^{28}$  \\                                          
  034411.22+320816.3	& 0.14  & M5.25	  & 209.3 &   13.02  &      III  &  6.38$\cdot10^{28}$  \\                                          
  034423.67+320646.5	& 0.39  &  M2.5	  & 222.7 &    9.83  &      III  &  4.42$\cdot10^{29}$  \\                                         
  034405.00+320953.8	& 0.93  &  K3.5	  & 129.5 &   21.37  &      III  &  2.99$\cdot10^{29}$  \\                                         
  034419.24+320734.7	& 0.26  & M3.75	  & 221.0 &    7.60  &       II  &  3.14$\cdot10^{29}$  \\                                          
  034416.43+320955.2	&  1.4  &    K0	  &  23.2 &    3.01  &      III  &  2.40$\cdot10^{30}$  \\   
  034421.56+321017.4	& 0.29  &  M1.5	  & 224.8 &    7.05  &   II/III  &  1.05$\cdot10^{30}$  \\   
  034421.91+321211.6	& 0.29  &    M4	  & 220.6 &   13.79  &      III  &  1.07$\cdot10^{29}$  \\ 
  034422.32+321200.8	& 0.33  &    M1	  & 228.4 &    8.42  &       II  &  1.66$\cdot10^{29}$  \\ 
  034422.98+321157.3	& 0.22  & M2.25	  & 225.9 &    5.09  &      III  &  1.00$\cdot10^{29}$  \\    
  034501.52+321051.5	& 0.94  &    K0	  &   --- &    1.89  &      III  &  7.60$\cdot10^{29}$  \\ 	 
  034425.58+321130.5	& 0.34  &    M0	  & 237.6 &     6.1  &      ---  &  8.99$\cdot10^{29}$  \\ 
  034426.69+320820.3	& 0.53  &  M0.5	  & 227.7 &    8.90  &      ---  &  3.81$\cdot10^{29}$  \\
  034433.31+320939.6	& 0.41  &    M2	  & 223.8 &    2.20  &   II/III  &  2.14$\cdot10^{29}$  \\
  034427.88+320731.6	& 0.46  &    M2	  & 222.5 &    5.97  &      III  &  2.15$\cdot10^{29}$  \\
  034432.77+320915.8	& 0.36  & M3.25	  & 221.7 &    5.39  &   II/III  &  4.38$\cdot10^{29}$  \\ 
  034432.58+320855.8	& 0.47  &    M3	  & 221.2 &    6.69  &   II/III  &  3.33$\cdot10^{29}$  \\
  034432.74+320837.5	&  4.4  &    G6	  &   8.1 &   2.64   &      III  &  4.84$\cdot10^{30}$  \\
  034433.98+320854.1	& 0.56  &    M0	  & 223.3 &   16.23  &      III  &  1.80$\cdot10^{30}$  \\
  034428.47+320722.4	& 0.71  &  K6.5	  & 251.3 &    7.02  &      III  &  6.82$\cdot10^{29}$  \\
  034441.31+321025.3	& 0.17  & M4.75	  & 216.0 &    3.72  &      III  &  2.86$\cdot10^{29}$  \\
  034439.21+320944.7	& 0.39  &    M2	  & 224.4 &    3.97  &       II  &  1.69$\cdot10^{30}$  \\
  034437.41+320900.9	& 0.51  &    M1	  & 225.8 &    8.39  &       II  &  5.75$\cdot10^{29}$  \\
  034442.58+321002.5	& 0.24  & M4.25	  & 219.7 &    3.56  &       II  &  2.46$\cdot10^{29}$  \\
  034438.70+320856.7	& 0.24  & M3.25	  & 222.3 &    2.09  &      III  &  2.70$\cdot10^{29}$  \\
  034438.72+320842.0	&  4.1  &    K3	  & 255.5 &    2.40  &      III  &  7.02$\cdot10^{30}$  \\
  034435.52+320804.5	& 0.09  & M5.25	  & 205.3 &    1.69  &      III  &  3.43$\cdot10^{28}$  \\
  034442.02+320900.1	& 0.37  & M4.25	  & 220.2 &   16.00  &      ---  &  2.84$\cdot10^{29}$  \\
  034437.89+320804.2	&  1.4  &    K7	  & 224.5 &    8.25  &       II  &  6.36$\cdot10^{30}$  \\
  034438.55+320800.7	& 0.72  & M1.25	  & 222.8 &    7.50  &       II  &  1.75$\cdot10^{30}$  \\
  034435.04+320736.9	&  1.5  &  K6.5	  & 221.3 &    4.50  &   II/III  &  2.53$\cdot10^{30}$  \\
  034438.47+320735.7	&  1.5  &    K6	  & 219.6 &    5.19  &       II  &  1.38$\cdot10^{30}$  \\
  034439.25+320735.5	&  3.6  &    K3	  & 218.6 &    5.20  &      III  &  5.65$\cdot10^{29}$  \\
  034443.53+320743.0	& 0.92  &    M1	  & 222.3 &    8.04  &      III  &  1.87$\cdot10^{30}$  \\
  034436.94+320645.4	&   17  &    G3	  &  16.1 &    1.68  &   II/III  &  9.32$\cdot10^{30}$  \\
  034434.88+320633.6	& 0.99  &  K5.5	  & 227.7 &    5.45  &      III  &  1.64$\cdot10^{30}$  \\
  034442.63+320619.5	& 0.19  &    M1	  & 229.2 &   11.80  &      III  &  3.51$\cdot10^{29}$  \\
  034437.41+320611.7	& 0.51  &    K7	  & 197.9 &    6.21  &      III  &  2.49$\cdot10^{29}$  \\
  034441.32+320453.5	&0.046  &    M5	  & 199.8 &    1.59  &      III  &  6.98$\cdot10^{28}$  \\
  034425.56+320617.0	& 0.54  & M2.25	  & 221.4 &    7.42  &       II  &  2.20$\cdot10^{30}$  \\   
  034427.02+320443.6	& 0.46  &    M1	  & 227.0 &    9.06  &      III  &  4.48$\cdot10^{30}$  \\
  034426.63+320358.3	& 0.39  & M4.75	  & 219.4 &    3.09  &      III  &  6.89$\cdot10^{29}$  \\   
  034426.03+320430.4	&  9.9  &    G8	  &  56.7 &    6.54  &       II  &  5.24$\cdot10^{30}$  \\
  034348.76+320733.4	& 0.34  &  M1.5	  & 226.2 &   20.30  &      III  &  3.41$\cdot10^{29}$  \\  
  034355.51+320932.5	&  1.9  &    K0	  &  60.0 &    4.86  &      III  &  1.85$\cdot10^{30}$  \\ 
  034349.39+321040.0	& 0.17  &  M3.5	  & 216.4 &   13.10  &      III  &  5.36$\cdot10^{29}$  \\   
  034423.99+321100.0	&  3.1  &    G0	  &   --- &    2.54  &      III  &  3.89$\cdot10^{29}$  \\ 	   
  034359.72+321403.2	& 0.33  & M0.75	  & 229.9 &   13.87  &      III  &  1.65$\cdot10^{31}$  \\    
  034404.25+321350.0	& 0.22  & M4.75	  & 217.6 &    1.20  &      III  &  7.33$\cdot10^{28}$  \\    
  034417.91+321220.4	& 0.31  &  M2.5	  & 223.5 &    4.47  &      III  &  6.08$\cdot10^{29}$  \\   
  034425.57+321230.0	& 0.45  &  M0.5	  & 230.3 &    8.38  &      III  &  1.97$\cdot10^{30}$  \\   
  034428.12+321600.3	&  0.3  & M3.25	  & 222.2 &    2.70  &      III  &  3.39$\cdot10^{29}$  \\    
  034439.81+321804.2	& 0.39  & M3.75	  & 220.8 &    9.50  &       II  &  1.62$\cdot10^{30}$  \\    
  034415.58+320921.9	&0.018  &  M7.5	  &   --- &    0.59   &      III  &  2.35$\cdot10^{28}$  \\  	
  034438.39+321259.8	& 0.21  &    M0	  & 235.2 &   13.49  &      III  &  1.76$\cdot10^{29}$  \\  
  034437.79+321218.2	& 0.19  &  M4.5	  & 217.9 &    3.27  &      III  &  3.99$\cdot10^{29}$  \\  
  034450.97+321609.6	& 0.35  & M3.25	  & 221.7 &   13.10  &      III  &  3.62$\cdot10^{29}$  \\   
  034440.13+321134.3	&  1.4  &    K2	  &  98.3 &    1.99  &      III  &  2.90$\cdot10^{30}$  \\
  034441.74+321202.4	& 0.17  &    M5	  & 214.4 &    2.10  &   II/III  &  1.28$\cdot10^{29}$  \\
  034448.83+321322.1	& 0.13  & M2.75	  & 225.2 &    6.90  &   II/III  &  2.67$\cdot10^{29}$  \\   
  034444.85+321105.8	& 0.17  & M2.75	  & 213.0 &    2.29  &      III  &  1.81$\cdot10^{29}$  \\   
  034501.74+321427.9	&  1.9  &    K4	  & 209.2 &   16.23  &      III  &  1.56$\cdot10^{30}$  \\   
  034450.65+321906.8	&  135  &    A0	  &   --- &    6.10  &      III  &  9.23$\cdot10^{30}$  \\   
  034507.61+321028.1	&  4.9  &    G1	  &   --- &    3.34  &      III  &  3.02$\cdot10^{29}$  \\    
  034455.63+320920.2	&  1.4  &    K4	  & 188.0 &    3.10  &      III  &  1.57$\cdot10^{30}$  \\   
  034456.15+320915.5	&  3.9  &    K0	  & 137.9 &    2.50  &   II/III  &  4.29$\cdot10^{30}$  \\   
  034453.76+320652.2	& 0.18  &    M4	  & 219.0 &    9.60  &      III  &  1.40$\cdot10^{29}$  \\   
  034456.12+320556.7	& 0.13  & M2.75	  & 225.2 &    3.30  &      III  &  6.61$\cdot10^{28}$  \\      
  034505.77+320308.2	& 0.54  &    M0	  & 221.4 &     4.9  &      ---  &  2.25$\cdot10^{30}$  \\   
  034418.26+320732.5	&0.078  & M4.75	  & 208.5 &    2.24  &   II/III  &  1.20$\cdot10^{29}$  \\      
  034423.57+320934.0	&0.074  &    M5	  & 205.1 &    1.72  &       II  &  3.74$\cdot10^{28}$  \\     
  034427.28+320717.6	&0.038  & M4.75	  & 202.2 &    1.53  &   II/III  &  6.62$\cdot10^{28}$  \\    
  034431.42+321129.4	&0.078  & M5.25	  & 203.0 &    1.40  &   II/III  &  1.28$\cdot10^{29}$  \\      
  034439.44+321008.2	&0.057  &    M5	  & 202.3 &    1.51  &      III  &  5.88$\cdot10^{28}$  \\     
  034351.24+321309.4	&  4.3  &    G5	  &   --- &    0.79  &      III  &  4.15$\cdot10^{30}$  \\     
  
\end{longtable}
}

\begin{table*}
\caption{Stellar parameters for IC~348 members with upper limits}
\label{tab:upper_limits}
\centering
\begin{tabular}{c c c c c c}
\hline\hline
\noalign{\smallskip}
Ra, Dec&$L_{\rm bol}$&Spectral Type&Turnover time&Period&X-ray Luminosity\\
 &$L_{\rm \odot}$&&days&days&erg/s \\
\noalign{\smallskip}
\hline
\noalign{\smallskip}

  034328.22+320159.1    & 0.768 &  M1.75  & 204.5 &      8.8 &  $<$1.67$\cdot10^{29}$  \\
  034345.16+320358.6    & 0.731 &     M0  & 222.2 &     27.6 &  $<$8.39$\cdot10^{28}$  \\  
  034359.08+321421.4    & 0.440 &   M3.5  & 212.3 &     17.7 &  $<$4.28$\cdot10^{28}$  \\                    
  034410.12+320404.4    &  0.12 &  M5.75  & 212.0 &      2.2 &  $<$6.41$\cdot10^{28}$  \\                                                      
  034418.19+320959.3    &  0.15 &  M4.25  & 212.5 &      2.7 &  $<$3.48$\cdot10^{28}$  \\
  034420.18+320856.7    &  0.32 &     M2  & 215.4 &      2.2 &  $<$1.79$\cdot10^{28}$  \\
  034421.26+321237.4    & 0.073 &  M4.75  & 213.4 &      2.3 &  $<$4.24$\cdot10^{28}$  \\
  034432.35+320327.3    & 0.028 &   M5.5  & 213.8 &      1.6 &  $<$3.72$\cdot10^{28}$  \\
  034432.80+320413.4    & 0.065 &     M5  & 213.4 &      1.3 &  $<$6.87$\cdot10^{28}$  \\
  034433.79+315830.3    & 0.492 &  M3.75  & 211.6 &      3.9 &  $<$6.32$\cdot10^{28}$  \\
  034434.05+320656.8    & 0.032 &  M7.25  & 138.3 &      3.4 &  $<$5.06$\cdot10^{28}$  \\
  034435.01+320857.4    & 0.032 &  M4.75  & 214.2 &      1.8 &  $<$2.16$\cdot10^{28}$  \\
  034435.44+320856.4    &  0.16 &  M5.25  & 211.5 &     1.98 &  $<$6.85$\cdot10^{28}$  \\
  034435.68+320303.6    &  0.13 &  M3.25  & 214.3 &     12.0 &  $<$4.17$\cdot10^{28}$  \\
  034436.98+320834.0    &  0.18 &  M4.75  & 211.5 &      2.5 &  $<$3.89$\cdot10^{28}$  \\
  034444.23+320847.4    & 0.040 &  M5.75  & 213.5 &      1.8 &  $<$3.84$\cdot10^{28}$  \\                                                                             
  034444.27+321036.8    & 0.017 &  M5.25  & 214.1 &      1.5 &  $<$3.61$\cdot10^{28}$  \\
  034445.67+321111.0    & 0.028 &  M4.75  & 214.3 &      1.3 &  $<$4.56$\cdot10^{28}$  \\

\hline
\end{tabular}
\end{table*}


\begin{thebibliography}{}

\bibitem[Bouvier et al.(1997)]{Bouvier97} Bouvier, J., Forestini, M., \& Allain, S.\ 1997, \aap, 326, 1023

\bibitem[Trujillo Bueno et al.(2004)]{Bueno04} Trujillo Bueno, 
J., Shchukina, N., \& Asensio Ramos, A.\ 2004, \nat, 430, 326 

\bibitem[Briggs et al.(2007)]{Briggs07} Briggs, K.~R., et al.\ 2007, \aap, 468, 413 

\bibitem[Broos et al.(2007)]{Broos07} Broos, P.~S. ,Feigelson, E.~D. ,Townsley, L.~K. ,Getman, K.~V., Wang, J.,Garmire, G.~P. ,Jiang, Z. ,Tsuboi, Y., 2007, \apjs, 169, 353 

\bibitem[Broos et al.(2010)]{Broos10} Broos, P.~S., Townsley, 
L.~K., Feigelson, E.~D., et al.\ 2010, \apj, 714, 1582   

\bibitem[Browning \& Basri(2007)]{Browning07} Browning, M.~K., \& Basri, G.\ 2007, 
Unsolved Problems in Stellar Physics, AIP Conference Proceedings, Volume 948,  157 

\bibitem[Chabrier \& K{\"u}ker(2006)]{Chabrier06} Chabrier, G., K{\"u}ker, M.\ 2006, \aap, 446, 1027 

\bibitem[Cieza 
\& Baliber(2006)]{Cieza06} Cieza, L., \& Baliber, N.\ 2006, \apj, 649, 862 

\bibitem[Cieza 
\& Baliber(2007)]{Cieza07} Cieza, L., \& Baliber, N.\ 2007, \apj, 671, 605

\bibitem[Cohen et al.(2004)]{Cohen04} Cohen, R.~E., Herbst, W., 
\& Williams, E.~C.\ 2004, \aj, 127, 1602 

\bibitem[Currie 
\& Kenyon(2009)]{Currie09} Currie, T., \& Kenyon, S.~J.\ 2009, \aj, 138, 703 

\bibitem[Dib et al.(2010)]{Dib10} Dib, S., Shadmehri, M., 
Padoan, P., Maheswar, G., Ojha, D.~K., 
\& Khajenabi, F.\ 2010, \mnras, 405, 401 

\bibitem[Dobler et al.(2006)]{Dobler06} Dobler, W., Stix, M., 
\& Brandenburg, A.\ 2006, \apj, 638, 336

\bibitem[Durney et al.(1993)]{Durney93} Durney, B.R., De Young, D.S.,
  Roxburgh, I.W.\ 1993, SolPhys, 145, 2070

\bibitem[Favata \& Micela(2003)]{FavataMicela03} Favata, F.~\&
Micela, G.\ 2003, Space Science Reviews, 108, 577

\bibitem[Feigelson \& Montmerle(1999)]{Feigelson99} Feigelson,
E.~D.~\& Montmerle, T.\ 1999, \araa, 37, 363

\bibitem[Getman et al.(2005)]{Getman05} Getman, K.~V., Flaccomio, E., Broos, P.S., et al.\ 2005, \apjs, 160, 319 

\bibitem[Getman et al.(2010)]{Getman10} Getman, K.~V., Feigelson, E.~D.,Broos, P.~S.,Townsley, L.~K.,Garmire, G.~P.\ 2010 \apj, 708, 1760

\bibitem[Giampapa et al.(1996)]{Giampapa96} Giampapa, M.S., Rosner, R., 
  Kashyap, V., Fleming, T.A.,  Schmitt, J.H.M.M., \& Bookbinder, J.A.\
   1996, \apj, 463, 707

\bibitem[Pietarila Graham et al.(2010)]{Graham10} Pietarila 
Graham, J., Cameron, R., \& Sch{\"u}ssler, M.\ 2010, \apj, 714, 1606 

\bibitem[G{\"u}del et al.(1997)]{Guedel97} G{\"u}del, M., Guinan,
E.~F., \& Skinner, S.~L.\ 1997, \apj, 483, 947

\bibitem[G{\"u}del et al.(2007)]{Guedel07} G{\"u}del, M., 
et al.\ 2007, \aap, 468, 353

\bibitem[Herbig(1954)]{Herbig54} Herbig, G.~H.\ 1954, \pasp, 66, 
19 

\bibitem[Herbig(1998)]{Herbig98} Herbig, G.~H.\ 1998, \apj, 497, 
736 

\bibitem[Herbst et al.(2000)]{Herbst2000} Herbst, W., Maley, 
J.~A., \& Williams, E.~C.\ 2000, \aj, 120, 349 

\bibitem[Herbst \& Mundt(2005)]{Herbst05} Herbst, W., \& Mundt, R.\ 2005, \apj, 633, 967 

\bibitem[Herbst et al.(2007)]{Herbst07} Herbst, W., Eisl{\"o}ffel, J., Mundt, R., 
\& Scholz, A.\ 2007, Protostars and Planets V, 297

\bibitem[Herbst(2008)]{Herbst08} Herbst, W.\ 2008, Handbook of 
Star Forming Regions, Volume I, 372 

\bibitem[I{\c s}{\i}k et al.(2011)]{Isik11} I{\c s}{\i}k, E., Schmitt, D., \& Sch{\"u}ssler, M.\ 2011, \aap, 528, A135

\bibitem[James et al.(2000)]{James00} James, D.~J., Jardine, 
M.~M., Jeffries, R.~D., Randich, S., Collier Cameron, A., 
\& Ferreira, M.\ 2000, \mnras, 318, 1217 

\bibitem[Jardine \& Unruh(1999)]{Jardine99} Jardine, M., \& Unruh, Y.~C.\ 1999, \aap, 346, 883 

\bibitem[Jeffries et al.(2011)]{Jeffries11} Jeffries, R.~D., 
Jackson, R.~J., Briggs, K.~R., Evans, P.~A., \& Pye, J.~P.\ 2011, \mnras, 411, 2099

\bibitem[Johns-Krull(2007)]{Johns-Krull07} Johns-Krull, C.~M.\ 2007, \apj, 664, 975 

\bibitem[Kim \& Demarque(1996)]{Kim96} 
Kim, Y.-C., \& Demarque, P.\ 1996, \apj, 457, 340

\bibitem[Kim et al.(1996)]{Kim_ea96} 
Kim, Y.-C., Fox, P.A., Demarque, P., \& Sofia, S.\ 1996, \apj, 461, 499

\bibitem[K{\i}z{\i}lo{\u g}lu et al.(2005)]{Kiziloglu05} 
K{\i}z{\i}lo{\u g}lu, {\"U}., K{\i}z{\i}lo{\u g}lu, N., 
\& Baykal, A.\ 2005, \aj, 130, 2766 

\bibitem[Kraft et al.(1991)]{Kraft91} 
Kraft, R.~P., Burrows, D.~N., Nousek, J.~A.\ 1991, \apj, 374, 344

\bibitem[K\"uker \& R\"udiger(1999)]{Kueker99} K\"uker, M., \& R\"udiger, G.\
1999, \aap, 346, 922

\bibitem[Lada 
\& Lada(1995)]{Lada95} Lada, E.~A., \& Lada, C.~J.\ 1995, \aj, 109, 1682 

\bibitem[Luhman et al.(1998)]{Lada98} Luhman, K.~L., Rieke, 
G.~H., Lada, C.~J., \& Lada, E.~A.\ 1998, \apj, 508, 347 

\bibitem[Lada et al.(2006)]{Lada06} Lada, C.~J., et al.\ 2006, 
\aj, 131, 1574 

\bibitem[Littlefair et al.(2005)]{Littlefair05} Littlefair, S.~P., 
Naylor, T., Burningham, B., \& Jeffries, R.~D.\ 2005, \mnras, 358, 341 

\bibitem[Luhman et al.(1998)]{Luhman98} Luhman, K.~L., Rieke, 
G.~H., Lada, C.~J., \& Lada, E.~A.\ 1998, \apj, 508, 347

\bibitem[Luhman(1999)]{Luhman99} Luhman, K.~L.\ 1999, \apj, 525, 466

\bibitem[Luhman et al.(2003)]{Luhman03} Luhman, K.~L., Stauffer, 
J.~R., Muench, A.~A., Rieke, G.~H., Lada, E.~A., Bouvier, J., 
\& Lada, C.~J.\ 2003, \apj, 593, 1093 

\bibitem[Luhman et al.(2005)]{Luhman05} Luhman, K.~L., Lada, 
E.~A., Muench, A.~A., \& Elston, R.~J.\ 2005, \apj, 618, 810

\bibitem[Luhman et al.(2010)]{Luhman10} Luhman, K.~L., Allen,
P.~R., Espaillat, C., Hartmann, L., \& Calvet, N.\ 2010, \apjs, 186, 111

\bibitem[Mayne et al.(2007)]{Mayne07} Mayne, N.~J., Naylor, T.,
Littlefair, S.~P., Saunders, E.~S.,
\& Jeffries, R.~D.\ 2007, \mnras, 375, 1220

\bibitem[Maggio et al.(1987)]{Maggio87} Maggio, A., Sciortino, S., 
Vaiana, G.S., et al.\ 1987, \apj, 315, 687

\bibitem[Messina et al.(2003)]{Messina03} 
Messina, S., Pizzolato, N., Guinan, E.F., \& and Rodono, M.\ 2003,
\aap, 410, 671

\bibitem[Montesinos et al.(2001)]{Montesinos01} Montesinos, B.,
 Thomas, J.H., Ventura, P.,~\& Mazzitelli, I.\ 2001, \mnras, 326, 877

\bibitem[Muench et al.(2003)]{Muench03} Muench, A.~A., Lada, E.~A., Lada, C.~J., Elston, R.~J., Alves, J.~F., Horrobin, M., Huard, T.~H., Levine, J.~L.,  Raines, S.~N., Rom{\'a}n-Z{\'u}{\~n}iga, C.\ 2003, \aj, 125, 2029

\bibitem[Muench et al.(2007)]{Muench07} Muench, A.~A., Lada, 
C.~J., Luhman, K.~L., Muzerolle, J., \& Young, E.\ 2007, \aj, 134, 411 

\bibitem[Nordhagen et al.(2006)]{Nordhagen06} Nordhagen, S., Herbst, W., Rhode, K.~L., Williams, E.~C. \ 2006, \aj, 132, 1555

\bibitem[Ossendrijver(2003)]{Ossendrijver03} Ossendrijver, M.\ 2003,
 \aapr, 11, 287

\bibitem[Pallavicini et al.(1981)]{Pallavicini81} Pallavicini, R., Golub, L., 
 Rosner, R., Vaiana, G.S., Ayres, T., \& Linsky, J.L.\ 1981, \apj 248, 279

\bibitem[Pizzolato et al.(2003)]{Pizzolato03} Pizzolato, N., Maggio, A.,
Micela, G., Sciortino, S., \& Ventura, P.\ 2003, \aap, 397, 147

\bibitem[Preibisch \& Zinnecker(2001)]{PZ01} Preibisch, T., \& Zinnecker, H.\ 2001, \aj, 122, 866 

\bibitem[Preibisch \& Zinnecker(2002)]{PZ02}
Preibisch, T., \& Zinnecker, H.\ 2002, \aj, 123, 1613

\bibitem[Preibisch et 
al.(2003)]{Preibisch03} Preibisch, T., Stanke, T., \& Zinnecker, H.\ 2003, \aap, 409, 147 

\bibitem[Preibisch \& Zinnecker(2004)]{PZ04} Preibisch, T., \& Zinnecker, H.\ 2004, \aap, 422, 1001

\bibitem[Preibisch \& Feigelson(2005)]{PF05} Preibisch, T., \& Feigelson, E.~D.\ 2005, \apjs, 160, 390

\bibitem[Preibisch, Zinnecker, \& Herbig(1996)]{Preibisch96} 
 Preibisch, T., Zinnecker, H., \& Herbig, G.H.\ 1996,  \aap, 310, 456

\bibitem[Preibisch et al.(2005)]{Preibisch05} Preibisch, T., Kim, Y.-C.,
Favata, F., et al.\ 2005, \apjs, 160, 401

\bibitem[Prosser et al.(1996)]{Prosser96} Prosser, C.~F., 
Randich, S., Stauffer, J.~R., Schmitt, J.~H.~M.~M., 
\& Simon, T.\ 1996, \aj, 112, 1570

\bibitem[Randich(1998)]{Randich98} Randich, S.\ 1998, in: Cool Stars, 
Stellar Systems, and the Sun, ASP Conf.~Ser.~154, p.~501 

\bibitem[Randich(2000)]{Randich00} Randich, S.\ 2000, Stellar 
Clusters and Associations: Convection, Rotation, and Dynamos, 
eds. R.~Pallavicini, G.~Micela, and S.~Sciortino,
ASP Conference Series Vol.~198, p.~401

\bibitem[Siess et 
al.(2000)]{Siess00} Siess, L., Dufour, E., \& Forestini, M.\ 2000, \aap, 358, 593 

\bibitem[Stauffer et al.(1997a)]{Stauffer97a} Stauffer, J.~R., 
Balachandran, S.~C., Krishnamurthi, A., Pinsonneault, M., Terndrup, D.~M., 
\& Stern, R.~A.\ 1997a, \apj, 475, 604 

\bibitem[Stauffer et al.(1997b)]{Stauffer97b} Stauffer, J.~R., 
Hartmann, L.~W., Prosser, C.~F., Randich, S., Balachandran, S., Patten, 
B.~M., Simon, T., \& Giampapa, M.\ 1997b, \apj, 479, 776 


\bibitem[Stelzer \& Neuh\"auser(2001)]{Stelzer01} Stelzer, B., Neuh\"auser, R.\ 2001, \aap, 377, 538

\bibitem[Stelzer et al.(2011)]{Stelzer11} Stelzer, B., Preibisch, 
T., Alexander, F., et al.\ 2011, arXiv:1111.4420 
 

\bibitem[Telleschi et al.(2005)]{Telleschi05} Telleschi, A.,
G{\"u}del, M., Briggs, K., Audard, M., Ness, J.-U.,
\& Skinner, S.~L.\ 2005, \apj, 622, 653

\bibitem[Townsley et al.(2003)]{Townsley03} Townsley, L.~K., Feigelson, E.~D., Montmerle, T., Broos, P.~S., Chu, Y.-H., Garmire, G.~P., \ 2003, \apj, 593, 874

\bibitem[V{\"o}gler \& Sch{\"u}ssler(2007)]{Voegler07} V{\"o}gler, A., \& Sch{\"u}ssler, M.\ 2007, \aap, 465, L43 

\bibitem[Weights et al.(2009)]{Weights09} Weights, D.~J., Lucas, 
P.~W., Roche, P.~F., Pinfield, D.~J., \& Riddick, F.\ 2009, \mnras, 392, 817 
\end{thebibliography}
\end{document}